\documentclass{aa}  

\usepackage[utf8]{inputenc}
\usepackage[T1]{fontenc}
\usepackage{multirow}
\usepackage{xcolor}
\usepackage{natbib}
\usepackage{mathtools}
\usepackage{hyperref}
\usepackage{graphicx}% Include figurfigure files
\usepackage{dcolumn}% Align table columns on decimal point
\usepackage{bm}% bold math
\usepackage{siunitx}					% SI units			
\usepackage{physics}
\usepackage{comment}
\DeclareSIUnit \parsec {pc}
\usepackage{txfonts}
\usepackage[switch]{lineno} 
\usepackage[normalem]{ulem}

\begin{document} 
% ==============================================================================================================
\title{3D dust attenuation effects in galaxies: structural biases and induced negative color gradients}

 %===============================================================================================================
   \author{
J.~Duarte \inst{1,}\thanks{joao.r.d.duarte@tecnico.ulisboa.pt},
S.~Gonz\'alez-Gait\'an \inst{2},
P.~Papaderos \inst{3},
A.~M.~Mour\~ao \inst{1},
and
R.~P.~Santos\inst{1.4}
          }

\authorrunning{J.~Duarte}

\institute{
CENTRA, Instituto Superior T\'ecnico, Universidade de Lisboa, Av. Rovisco Pais 1, 1049-001 Lisboa, Portugal
\and
Instituto de Astrof\'isica e Ci\^encias do Espaço, Universidade de Lisboa – OAL, Tapada da Ajuda, 1349-018 Lisboa, Portugal
\and
Instituto de Astrof\'isica e Ci\^encias do Espaço, Universidade do Porto – CAUP, Rua das Estrelas, 4150-762 Porto, Portugal
\and
European Southern Observatory, Alonso de C\'ordova 3107, Casilla 19, Santiago, Chile
}
\date{Accepted XXX/ Received November 2026}
\abstract{Radial surface brightness profiles (SBPs) and color gradients are particularly useful observational tools for studying galaxy structure, formation, and evolution. While it is usually assumed that these observables are primarily driven by stellar properties, such as mass, age or metallicity, their dependence on three-dimensional (3D) dust attenuation is not fully understood or accounted for.}
{We analyze how 3D dust attenuation affects galaxy observations by obscuring stellar populations and hindering the estimation of intrinsic properties. We examine dust-induced shifts in surface brightness and color for a variety of galaxy types and dust masses.}
{Using SKIRT radiative transfer models, we simulated simple Sérsic galaxies, with spherically-symmetric stellar densities, constant stellar-to-dust ratios, and a spatially constant spectral energy distribution (SED). We fitted and compared the attenuated and dust-free projected SBPs. We also computed dust-induced color gradients and compared them to those predicted by a foreground dust-screen.}
{We find that the attenuated SBPs differ significantly from the dust-free projected profiles, particularly in the central galactic regions. These differences become more pronounced at bluer wavelengths and for higher core dust densities, and translate to variations in the fitted Sérsic index, $n$, of up to $\sim25\%$. This can lead to incorrect conclusions about the intrinsic stellar distribution and mimic a central mass deficit. For galaxies with very centrally-dense dust distributions, a core-Sérsic model results in a much better fit of the attenuated SBP, as indicated by the Akaike information criterion, and it can be used to recover the original simulation parameters, in contrast to a standard Sérsic fit. We also find that, for dust masses as low as $\sim2\times10^6 M_{\odot}$, dust-induced negative color gradients reach the same order of magnitude as those commonly observed in galaxies ($\sim -0.1$ mag/dex), which a dust-screen fails to account for. This occurs even when assuming a spatially invariant SED and can lead to wrong conclusions about intrinsic stellar populations and related astrophysical questions, such as their formation histories or the mechanisms of inside-out star formation quenching.}
{A full 3D approach is necessary to properly model dust attenuation effects in galaxy observations. These effects, if not accounted for, can bias inferences made about stellar populations, their evolutionary histories, and spatial distributions.}

% 5 {} token are mandatory

\maketitle
%_____________________________________________________________

\section{Introduction}
 \label{sec:intro}
Radial surface brightness profiles (SBPs) and color gradients are powerful observational tools for the study of galaxies. SBPs encode vital information about stellar structure \citep[e.g.,][]{der_Wel_2012}, while color gradients are crucial to our understanding of galaxy formation and evolution \citep[e.g.,][]{Tamura00}.
\par
A common description of the observed SBP for many stellar spheroids, particularly ellipticals and spiral bulges, is given by the \cite{Sersic63} law, as defined by Eq. \ref{eq:sersic_profile}:

\begin{equation}
    I_{\textrm{S}}(r)=I_0 \exp{-b_n \left[ \left(\frac{r}{r_{\textrm{eff}}} \right)^{1/n} -1 \right]},
    \label{eq:sersic_profile}
\end{equation}

\noindent
where $n$ is the Sérsic index, $r_{\mathrm{eff}}$ is the effective radius, $I_0$ is a normalization constant, and $b_n$ is defined so that half the total luminosity is contained within $r_{\mathrm{eff}}$. The Sérsic index is often used to characterize galaxies \citep[e.g.,][]{der_Wel_2012}. For instance, $n=1$ is characteristic of dwarf ellipticals \citep{Binggeli_1991}, while $n=4$ corresponds to the \cite{de_Vaucouleurs_1948} law, which gives a good approximation to massive early-type galaxies (ETGs), and $n=2$ roughly coincides with the division between classical and pseudo-bulges \citep[e.g.,][]{Gadotti09,Neumann17}. However, the extent to which observed SBPs are representative of intrinsic stellar distributions is largely unexplored, as it is not clear how 3D dust attenuation affects them.

\par
Radial color gradients are also well documented \citep[e.g.][]{Peletier90a,de_jong1996}. Early surface photometry of ETGs showed systematic negative gradients in the optical of the order of a few tenths of a mag over the galaxy extent \citep[e.g.,][]{Vader88,Franx_1990,Tamura00,denBrok11}. While these gradients are rather weak, with $\nabla_{U-R}\sim -0.23$ mag/dex and $\nabla_{B-R}\sim -0.1$ mag/dex \citep[e.g.,][]{Peletier90a}, they become stronger for optical/near-infrared (NIR) data, up to $\nabla_{V-K}\sim -0.6$ mag/dex \citep{Peletier90b}. The origin of these gradients is commonly attributed to radial variations in stellar age or metallicity \citep[e.g.,][]{Tamura00,Jablonka07,LaBarbera2009,Lin_2024}. However, dust attenuation can produce similar color variations \citep[e.g.,][]{Duarte_2025}, making it difficult to differentiate between intrinsic and dust-induced gradients. 

\par
Traditionally, SBPs and color indices are translated into mass, age or metallicity through comparison with predictions from evolutionary stellar synthesis models. These models are often zero-dimensional, assuming that the integrated spectral energy distribution (SED) of a galaxy behaves like a point-source, with dust attenuation approximated by a foreground screen \citep[e.g.,][]{Martorano_2026}. Such models do not account for possible effects arising from dust geometry \citep{Duarte_2025}. For example, given that the SED recorded at the center of a galaxy is subject to higher levels of dust attenuation than in the outskirts, it is expected that dust would lead to systematic radial changes to both the observed SBP and color gradients.
%, 
\par
While one can account for varying column densities by fitting for individual pixel-by-pixel dust reddening, this only provides a partial solution. Full radiative transfer is required to properly model the realistic 3D mixture of stars and dust, given that the SED recorded at a given projected radius $r$ is not identical to that of a simple or even composite stellar population (SSP and CSP, respectively), but rather depends on the complex interplay between the dust and star geometries, where absorption and scattering, both out and into the line-of-sight, play significant roles.
\par
We aim to investigate the extent to which observed SBPs and color gradients are affected by dust attenuation. We are interested in whether dust can significantly alter an observed Sérsic profile and the degree to which dust-induced effects are relevant in color gradient determinations (i.e., larger than typical uncertainties of the order of $\sim$0.1 mag/dex \citep[e.g.,][]{Miller_2023}). To this end we use the radiative transfer code SKIRT \citep[][]{Camps_2015,CAMPS_2020} to model a set of simplified galaxy geometries. In Sect.~\ref{sec:methods}, we describe SKIRT and outline the proposed models and their assumptions. The results are presented in Sect.~\ref{sec:results} and our main findings are discussed and summarized in Sect.~\ref{sec:summary}.

\section{Methods}
\label{sec:methods}
\subsection{Galaxy simulations}
\label{sec:methods_sim}
In this work, we used the radiative transfer code SKIRT \citep{Camps_2015,CAMPS_2020}\footnote{\url{https://skirt.ugent.be/root/_home.html}} to simulate observations of galaxies by tracing the path of simulated photons as they traverse a 3D dust medium, accounting for both scattering and absorption. In this way, we investigated dust-induced shifts in surface brightness and color across the full projected galaxy. We adopted a simple experimental setup, consisting of a spherically-symmetric stellar system with sources of constant age and metallicity, and thus a constant SED. The source system is homogeneously mixed with dust, such that the dust-to-stellar mass ratio is spatially invariant.

\par
The stellar SED was modeled by a \citet{Bruzual_2003} SSP, assuming a \cite{chabrier_2003} initial mass function, a metallicity of $0.02$ and an age of $5$ Gyr, roughly corresponding to an ETG at cosmic noon. We assumed a spherically-symmetric intrinsic stellar distribution which, in projection, leads to a \cite{Sersic63} profile, normalized so the total luminosity of the galaxy was $10^{10}L_{\odot}$. An effective projected radius of $r_{\rm eff, sim}=4000$ pc was used and the Sérsic index was taken to be $n_{\textrm{sim}}=1$, 2 or 4, corresponding to the different galaxy types detailed in Section \ref{sec:intro}.

\par
A \cite{Zubko_2004} dust model was assumed, with a spatially invariant stellar-to-dust ratio. The dust density distribution was taken to be the same as the stellar distribution, normalized by the total dust mass of the galaxy $M_D$, with observations simulated for different values of $M_D$. Dust emission was neglected.

\par
For each model, SKIRT's ``OligoExtinctionOnly'' mode was used to simulate the path of $10^{11}$ photon packets for wavelengths corresponding to some common photometric filters, namely the $U$, $B$, $V$, $R$ and $I$ bands. While observations in these filters shape much of our understanding of galaxies, the dust effects to which they are subject are often underestimated and corrected for over-simplistically via a foreground dust screen model. Each galaxy was simulated at a distance of $10$ Mpc with a field of view of $10$ kpc and a pixel size of $25$ pc. Point spread function convolution was not taken into account, as our goal is not to generate realistic observations, but to quantify dust effects. From the simulations, we generated two sets of SBPs: ``attenuated'' profiles, for which the full dust effects were taken into account, and ``transparent'' profiles, which are dust-free.

\subsection{Surface brightness fits and color gradients}
\label{sec:methods_fits}
We used the astropy package \citep{astropy} and a Bayesian framework to fit 1D Sérsic profiles, as described by Eq. \ref{eq:sersic_profile}, to both attenuated and transparent SBPs, comparing the best-fit parameters to the simulation input. For the $n_{\textrm{sim}}=4$ model, we also fitted a core-Sérsic profile \citep{Graham_2003,Trujillo_2004} to the attenuated SBP, which usually describes ETGs with depleted stellar cores and is given by Eq. \ref{eq:sersic_core_profile}:

\begin{equation}
    I_{\textrm{SC}}(r)=I'_0 \left[ 1+ \left(\frac{r_b}{r}\right)^{\alpha}\right]^{\frac{\gamma}{\alpha}} \exp{-b_n \left[ \left(\frac{r^{\alpha}+r_b^{\alpha}}{r_{\textrm{eff}}^{\alpha}} \right)^{\frac{1}{\alpha n}} -1 \right]}.
    \label{eq:sersic_core_profile}
\end{equation}
\noindent
For small $r$, the core-Sérsic behaves like a power law with negative logarithmic slope, $\gamma$, while for large $r$ it resembles a Sérsic profile. The break radius, $r_b$, separates both components and $\alpha$ controls the transition sharpness. $I'_0$ is a normalization constant.

\par
From the simulated attenuated profiles for a given color $c$, we computed the dust-induced color gradient, $\nabla_c$, as given by Eq. \ref{eq:c_grad}:

\begin{equation}
\nabla_c =\frac{\Delta c}{\Delta\log(r)}=\frac{c\left(r_{\rm eff, sim}\right)-c\left(0.1 r_{\rm eff, sim}\right)}{\log(r_{\rm eff, sim}) - \log(0.1 r_{\rm eff, sim})},
    \label{eq:c_grad}
\end{equation}
\noindent
where $c(r)$ is the mean color of all pixels at projected radius $r$. 

\section{Results}
\label{sec:results}

\subsection{Surface brightness profiles}
\label{sec:lum_profile}

\begin{figure*}
	\centering
	\includegraphics[width=.9\textwidth]{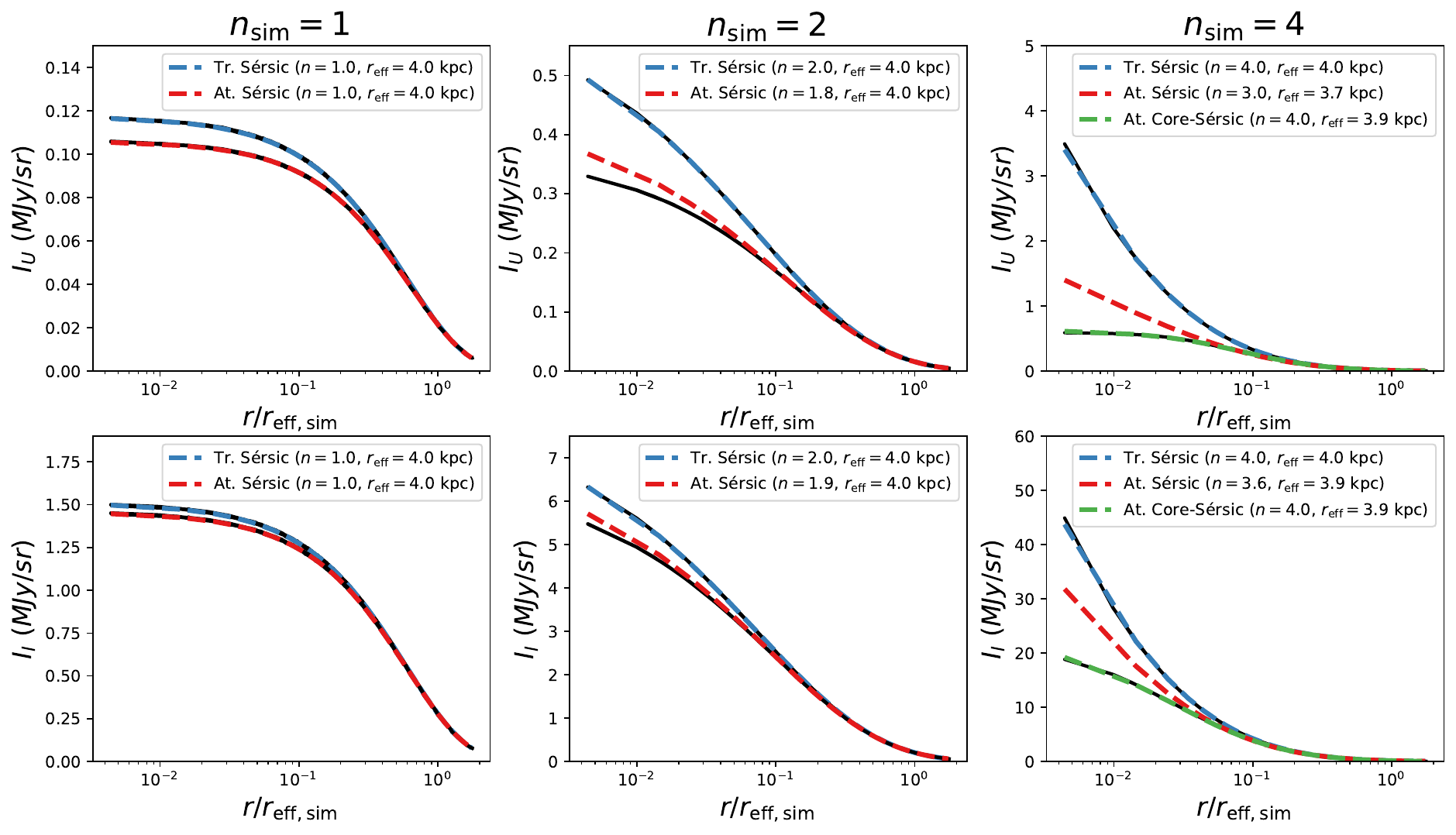}
	\caption{Transparent and attenuated SBPs for simulated Sérsic galaxies with $n_{\textrm{sim}}=1$ (left), $n_{\textrm{sim}}=2$ (center), and $n_{\textrm{sim}}=4$ (right), in both $U$ (top) and $I$ (bottom) bands, assuming $r_{\textrm{eff,sim}}=4$ kpc and $M_{\mathrm{D}}=10^6 M_{\odot}$. Sérsic best-fit curves to the transparent and attenuated SBPs are plotted in blue and red, respectively. Core-Sérsic best-fit curves for the $n_{\textrm{sim}}=4$ attenuated SBPs are plotted in green. Best-fit parameters are also shown.}

\label{fig:lum_profiles}
\end{figure*}

Simulated transparent and attenuated SBPs in the $U$ and $I$ bands are shown in Fig. \ref{fig:lum_profiles}, for a dust mass $M_{\mathrm{D}}=10^6 M_{\odot}$. Best-fit curves for the Sérsic and core-Sérsic models are also shown. Best-fit parameters and Akaike information criterion (AIC) values for the two models are found in Tabs. \ref{tab:sbp_params_s} and \ref{tab:sbp_params_sc}, respectively.

\par
We find that the transparent SBP fit can recover the simulation profile input parameters, with relative errors of $<0.1\%$ in $n$ and $<0.6\%$ in $r_{\rm eff}$. The presence of dust significantly impacts the attenuated SBPs, which diverge from the dust-free profiles, particularly in the central regions of the galaxy. This divergence is most significant at bluer wavelengths and tends to increase with the Sérsic index. While $r_{\rm eff}$ is mostly unaffected, relative errors in $n$ can be as large as $\sim25\%$, in the case of the $U$-band SBP fit for the $n_{\textrm{sim}}=4$ galaxy, for which we recovered $n=2.957_{-0.009}^{+0.007}$. We also find that the fitted $n$ decreases as the dust mass increases (see Appendix \ref{app:sbp_mass}).

\par
For higher Sérsic indices, we find that the shape of the SBPs is noticeably altered, departing from the standard Sérsic law, especially in the central regions. In particular, for the $n_{\textrm{sim}}=4$ galaxy, we find that the attenuated SBPs are better described by a core-Sérsic model, as confirmed by the AIC values in Tab. \ref{tab:sbp_params_sc}. We also note that, when using a core-Sérsic profile to fit the attenuated SBP, we are able to recover the original simulation input Sérsic indices, in contrast to the standard Sérsic law fit. This is expected, as, in the core-Sérsic case, $n$ is set by the outer SBP.

\par
Test simulations for a modified ETG, with a projected Sérsic stellar profile with $n_{\textrm{sim}}=4$ and a less centrally-dense \cite{Plummer_1911} dust density distribution, result in attenuated SBPs which, while diverging from the transparent SBPs, can be accurately described by the standard Sérsic law. This shows that the divergence between the attenuated and transparent SBPs is primarily driven by the dust distribution, rather than the stellar profile. A more detailed explanation of the dust effects can be found in \citep{Duarte_2025} and in Appendix \ref{app:flux_comp}, but, in general, larger deviations are linked to galaxies with higher core dust densities. 

\subsection{Color gradients}
\label{sec:color_grad}
\begin{figure*}
	\centering
	\includegraphics[width=\textwidth]{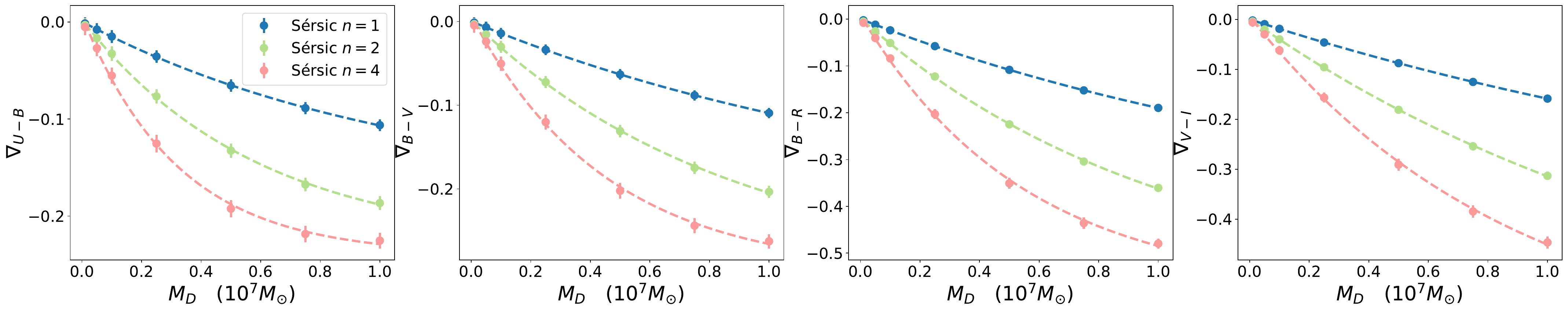}
	\caption{Color gradients for $U-B$, $B-V$, $B-R$ and $V-I$ as a function of the dust mass $M_D$. Results for three galaxies with separate Sérsic profiles are shown, with index $n_{\textrm{sim}}=1$ (blue), $n_{\textrm{sim}}=2$ (green) and $n_{\textrm{sim}}=4$ (red). The dashed lines correspond to exponential fits of the data.}

\label{fig:grads}
\end{figure*}

Simulated radial color gradients for $U-B$, $B-V$, $B-R$ and $V-I$ are plotted in Fig. \ref{fig:grads} for different galaxy types and dust masses, along with best-fit exponential curves. Additional results for $V-K$ are shown in Appendix \ref{app:k-band}. We note that, while simulated colors are influenced by the SSP, induced gradients only depend on dust mass and star/dust geometry (see Appendix \ref{app:gal_age}).

\par
We find that dust-induced gradients are negligible for small dust masses, but grow as dust mass increases. Dust distributions have a large impact on the colors, with higher values of $n_{\textrm{sim}}$ resulting in larger gradients. For galaxies with higher core dust densities ($n_{\textrm{sim}}=2$ and $n_{\textrm{sim}}=4$) and dust masses as low as $\sim2\times10^6 M_{\odot}$, these gradients become larger than the typical uncertainties of $\sim0.1$ mag/dex, reaching the same order of magnitude as observed gradients \citep[e.g.,][]{Peletier90a,Breda_phd_2019}. This happens due to the different levels of attenuation at different radii, even in the absence of intrinsic SED variation.

\par
\begin{figure*}
	\centering
	\includegraphics[width=0.9\textwidth]{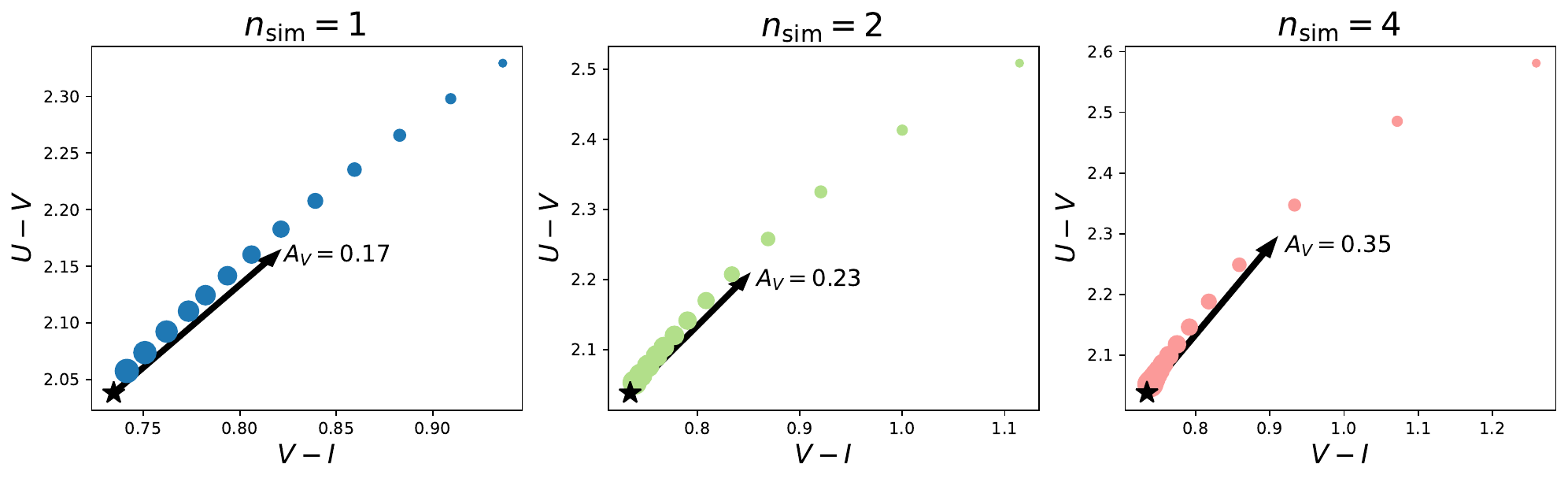}
	\caption{$U-V$/$V-I$ color-color diagram for simulated Sérsic galaxies with $M_{\mathrm{D}}=10^7 M_{\odot}$ in annuli with radii increasing in increments of $0.1 r_{\rm eff,sim}$. Larger annuli correspond to larger points on the plot. Results for the $n_{\textrm{sim}}=1$ (blue), $n_{\textrm{sim}}=2$ (green) and $n_{\textrm{sim}}=4$ (red) profiles are shown. The transparent color values are plotted as a black star and global $A_V$ vectors induced by a dust screen are indicated by black arrows.}

\label{fig:_uv-vi}
\end{figure*}

In Fig. \ref{fig:_uv-vi}, we plot the $U-V$/$V-I$ color-color diagram for different galactocentric annuli, with radii increasing in increments of $0.1 r_{\rm eff,sim}$. Color values for the transparent SSP are also shown, as well as the color vectors that would be induced by a dust screen with $R_V=3.068$, as defined by the \cite{Zubko_2004} mix, and $A_V$ set to the global value measured for each galaxy. As an example, $M_{\mathrm{D}}=10^7 M_{\odot}$ was assumed, but we found similar results across all dust masses. We find that a dust screen approach can not account for the full dust-induced gradients, even for fairly negligible $A_V$ values. We also note that global $A_V$ values are greatly influenced by galactic geometry, making them less than ideal metrics for dust correction, even in the best-case scenario where both transparent and attenuated SEDs are known.

\section{Discussion and conclusions}
\label{sec:summary}

Using radiative transfer simulations, we have shown that attenuation can significantly alter the shape of observed SBPs, particularly for galaxies with centrally dense dust distributions. As a baseline, for $n_{\rm sim}=4$ ETGs, it can reduce the observed Sérsic index by as much as $25\%$. We have also found that the fitted $n$ decreases as the dust mass increases, meaning that, for very dusty high-mass ETGs, $n$ can be underestimated by up to $68\%$.

\par
The attenuated SBPs of high-$n$ ETGs also exhibit a central intensity flattening within $\sim0.1 r_{\rm eff,sim}$, which can be well described by a core-Sérsic model. This model is usually associated with a core mass deficit, as a consequence of gas-poor galaxy mergers in which the resulting supermassive black hole binary causes the ejection of stellar material \citep{Merritt_2006,Gualandris_2008}. Our modeling shows, however, that such profiles can also arise from radiative transfer effects, which are significant even for a modest amount of dust ($M_{\mathrm{D}}=10^6 M_{\odot}$), well within the range observed for ETGs \citep{Smith_2012}. Fitting the attenuated SBP with a core-Sérsic model allows the recovery of the dust-free projected profile Sérsic index, which is useful for systematizing the structural properties of stellar spheroids. However, it does not itself reveal whether the central flattening arises from dust effects or core mass deficit, meaning that other tracers, such as radio or X-ray data, are needed to distinguish between the two scenarios \citep[e.g.,][]{Krajnovic_2013}.
\par
Another central result is that attenuation in a dusty medium produces negative color gradients, even when the stellar SED and the dust-to-stellar mass ratio are spatially uniform. These induced gradients exceed typical observational uncertainties and fall well within the range of observed values for galaxy bulges and ETGs ($\sim0.1$ mag/dex). For ETGs, whose mean dust-to-stellar mass ratio is $\log(M_{\rm D}/M_{\star})=-4.42$ \citep{Smith_2012}, such gradients could arise for stellar masses $M_{\star}\sim5\times10^{10}M_{\odot}$, even for dust distributions with relatively low central density, such as the \cite{Plummer_1911} profile. We also note that a dust screen approach fails to account for these induced gradients.

\par
The fact that negative color gradients and changes to the shape of projected SBPs can arise from attenuation effects is important for many astrophysical questions, such as the structure and formation history of galaxy bulges, bars, and disks, and the mechanisms of inside-out star formation quenching \citep{Gadotti09,Perez_2013,breda_2020,Polis_2022}. Future work that improves the fidelity of simulated galaxies, with more realistic stellar and dust distributions, is necessary for the correct interpretation of galaxy observations.

\begin{acknowledgements}
This work was supported by FCT through grants No. UID/PRR/00099/2025 (DOI 10.54499/UID/PRR/00099/2025), UID/00099/2025 (DOI 10.54499/UID/00099/2025) and UID/04434/2025 (DOI 10.54499/UID/04434/2025). JD and RPS acknowledge FCT support under PhD grants 2023.01333.BD (DOI 10.54499/2023.01333.BD) and  2024.03599.BD (DOI 10.54499/2024.03599.BD), respectively. PP acknowledges FCT support through grant 2024.17990.PEX, and PI contract CIAAUP-092023-CTTI. JD thanks Maarten Baes for helpful discussions about SKIRT.

\end{acknowledgements}
% ::::::::::::::::::::
\bibliographystyle{aa} % style aa.bst
\bibliography{aanda.bib}

\newpage

\begin{appendix}
\onecolumn
\section{Surface brightness profile fits}
Best-fit parameters and AIC values for the Sérsic and core-Sérsic fits discussed in Section \ref{sec:lum_profile} are displayed in Tabs. \ref{tab:sbp_params_s} and \ref{tab:sbp_params_sc}, respectively.

\renewcommand{\arraystretch}{1.5}
\begin{table}[h]
    \centering
        \caption{Best-fit parameters and AIC values for the SBP fits using a Sérsic model, as described in Section \ref{sec:lum_profile}.}
    \begin{tabular}{c c c c c c c}
         Simulated Profile & Band & Type & $I_0$ (MJy/sr) &  $n$ & $r_{\textrm{eff}}$ (pc) & AIC  \\ \hline
         
         \multirow{4}{*}{$n_{\textrm{sim}}=1$} &  \multirow{2}{*}{$U$} & Transparent & $\left(2.191_{-0.004}^{+0.004}\right)\times10^{-2}$& $1.000_{-0.002}^{+0.002}$ & $3978_{-4}^{+4}$ & 1925 \\
         & & Attenuated & $\left(2.124_{-0.007}^{+0.003}\right)\times10^{-2}$& $0.965_{-0.002}^{+0.002}$ & $4024_{-3}^{+6}$ & 1869 \\
         &  \multirow{2}{*}{$I$} & Transparent & $\left(2.817_{-0.004}^{+0.004}\right)\times10^{-1}$& $1.000_{-0.002}^{+0.002}$ & $3979_{-3}^{+3}$ & 1918 \\
         & & Attenuated & $\left(2.787_{-0.005}^{+0.004}\right)\times10^{-1}$& $0.988_{-0.002}^{+0.002}$ & $3995_{-3}^{+4}$ & 1898 \\\hline

         \multirow{4}{*}{$n_{\textrm{sim}}=2$} &  \multirow{2}{*}{$U$} & Transparent & $\left(1.601_{-0.005}^{+0.006}\right)\times10^{-2}$& $1.999_{-0.005}^{+0.005}$ & $3989_{-6}^{+8}$ & 1291 \\
         &  & Attenuated & $\left(1.586_{-0.010}^{+0.004}\right)\times10^{-2}$& $1.819_{-0.004}^{+0.056}$ & $3997_{-6}^{+7}$ & 1431 \\
         &  \multirow{2}{*}{$I$} & Transparent & $\left(2.057_{-0.006}^{+0.005}\right)\times10^{-1}$& $2.000_{-0.004}^{+0.004}$ & $3990_{-5}^{+6}$ & 1284 \\
         &  & Attenuated & $\left(2.052_{-0.006}^{+0.005}\right)\times10^{-1}$& $1.934_{-0.004}^{+0.005}$ & $3992_{-5}^{+5}$ & 1289 \\\hline

         \multirow{4}{*}{$n_{\textrm{sim}}=4$} &  \multirow{2}{*}{$U$} & Transparent & $\left(1.153_{-0.009}^{+0.2}\right)\times10^{-2}$& $4.00_{-0.02}^{+0.02}$ & $3989_{-448}^{+13}$ & 1032 \\
         &  & Attenuated & $\left(1.311_{-0.005}^{+0.005}\right)\times10^{-2}$& $2.957_{-0.009}^{+0.007}$ & $3740_{-8}^{+7}$ & 5744 \\
          &\multirow{2}{*}{$I$} & Transparent & $\left(1.480_{-0.007}^{+0.007}\right)\times10^{-1}$& $4.00_{-0.01}^{+0.01}$ & $3993_{-10}^{+10}$ & 1033 \\
          &  & Attenuated & $\left(1.566_{-0.008}^{+0.006}\right)\times10^{-1}$& $3.58_{-0.01}^{+0.01}$ & $3881_{-8}^{+10}$ & 1754 \\ \hline
         
    \end{tabular}
    \tablefoot{Fit results for both the transparent and attenuated SBPs in the $U$ and $I$ band are listed. The simulated profiles were generated assuming an effective radius $r_{\textrm{eff,sim}}=4000$ pc. Errors on the parameters correspond to a 68\% credible region.}
    \label{tab:sbp_params_s}
\end{table}

\begin{table}[h]
    \centering
        \caption{Same as Tab. \ref{tab:sbp_params_s}, but for the core-Sérsic model.}
    \begin{tabular}{c c c c c c c c c c}
         Simulated Profile & Band & Type & $I'_0$ (MJy/sr) &  $n$ & $r_{\textrm{eff}}$ (pc) & $\gamma$ & $r_b$ (pc) & $\alpha$& AIC  \\ \hline
         \multirow{2}{*}{$n_{\textrm{sim}}=4$} & $U$ & Attenuated & $24_{-2}^{+2}$& $3.97_{-0.05}^{+0.04}$ & $3908_{-19}^{+19}$ & $0.02_{-0.01}^{+0.02}$ & $230_{-6}^{+8}$ & $1.52_{-0.04}^{+0.05}$ & 1076 \\
         & $I$ & Attenuated & $314_{-13}^{+19}$& $3.98_{-0.03}^{+0.03}$ & $3940_{-24}^{+19}$ & $0.11_{-0.10}^{+0.07}$ & $89_{-22}^{+17}$ & $1.3_{-0.1}^{+0.1}$ & 1727 \\\hline

    \end{tabular}

    \label{tab:sbp_params_sc}
\end{table}

\section{Impact of dust mass on the observed surface brightness profiles}
\label{app:sbp_mass}
In Section \ref{sec:lum_profile}, we explored the role that galactic dust can have on the observed SBPs for a set of galaxies with dust mass $M_{\mathrm{D}}=10^6 M_{\odot}$. We focused primarily on the effects arising from different dust distributions. In this Appendix, we take a closer look at the role of dust mass and its influences on the attenuated profiles.

\begin{figure}[!h]

	\includegraphics[width=0.9\textwidth]{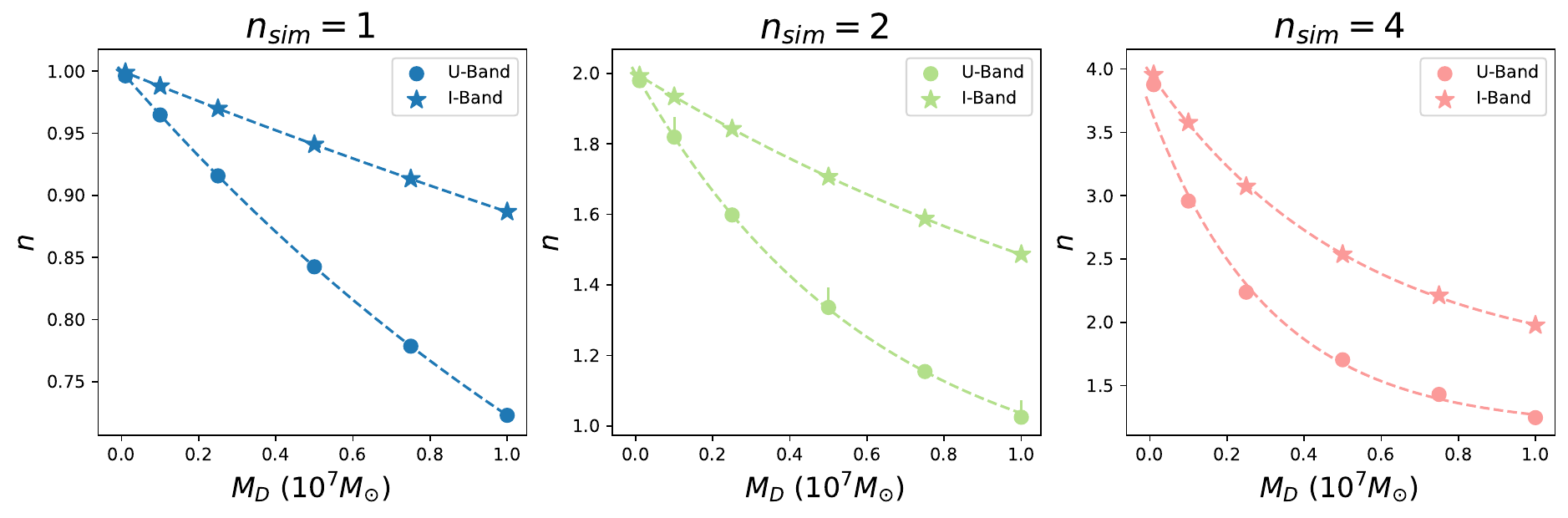}
	\caption{Best-fit Sérsic indices $n$ for the $U$- (circles) and $I$-band (stars) attenuated SBP as a function of dust mass $M_D$ for simulated Sérsic galaxies with $n_{\textrm{sim}}=1$ (left), $n_{\textrm{sim}}=2$ (center), and $n_{\textrm{sim}}=4$ (right). Errors in $n$ correspond to a 68\% credible region. The dashed lines correspond to exponential fits of the data.}

\label{fig:mass_vs_n}
\end{figure}
\par
For the three galaxy types discussed in Section \ref{sec:methods_sim}, the best-fit Sérsic indices $n$ for the $U$- and $I$-band attenuated SBP are plotted in Fig. \ref{fig:mass_vs_n} as a function of dust mass $M_D$. We find that the divergence between $n_{\textrm{sim}}$ and $n$ grows with the amount of dust, with $n$ decreasing exponentially with the increase in $M_D$, which results in overall shallower observed SBPs across all tested dust density distributions.
\par
Despite the fact that the dust-to-stellar ratio for both ETGs and dwarf elliptical galaxies is not constant, it has been generally reported that dust mass increases with stellar mass \citep{Garn_2010,Alighieri_2013}. It has also been reported that the observed Sérsic index increases with galaxy luminosity and stellar mass \citep[e.g.,][]{Young_1994,Binggeli_1998}. It is evident from Fig. \ref{fig:mass_vs_n} that this increase cannot be explained by dust attenuation effects and therefore likely reflects a real intrinsic trend. However, it is likely that radiative transfer effects are responsible for systematically lowering the Sérsic indices measured for high-mass galaxies. For example, for a high-mass, dusty ETG with $n_{\rm sim}=4$, the observed $U$-band attenuated Sérsic index can be underestimated by up to $\sim 68\%$. Thus, we expect the mass-$n$ relation to be more pronounced than commonly reported.

\section{Dust attenuation mechanisms}
\label{app:flux_comp}

\begin{figure}[!h]
\centering
	\includegraphics[width=0.5\textwidth]{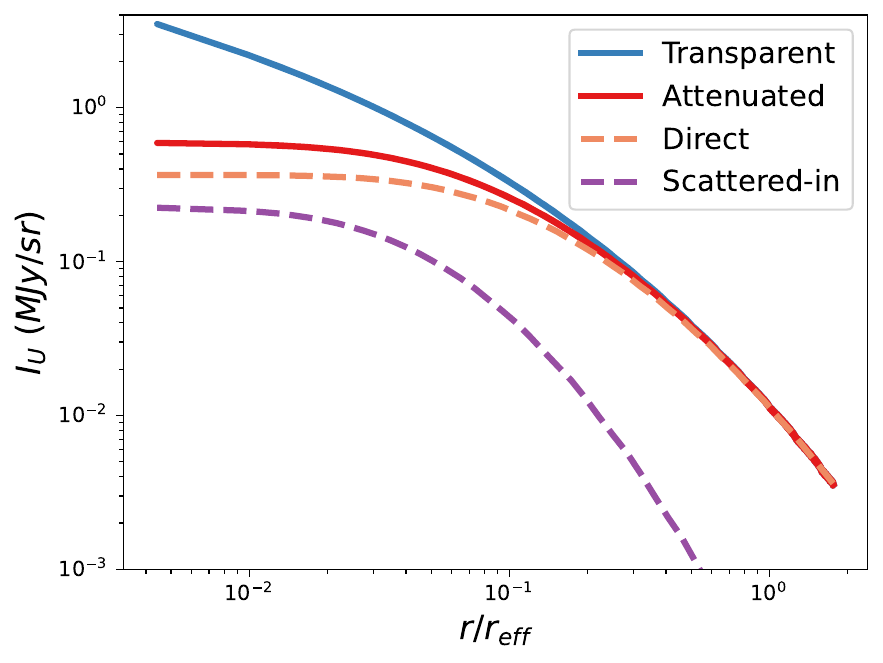}
	\caption{SBPs for a simulated Sérsic galaxy with $n_{\textrm{sim}}=4$, $r_{\textrm{eff,sim}}=4000$ pc and $M_{\mathrm{D}}=10^6 M_{\odot}$. The transparent profile is shown in blue, while the attenuated profile is shown in red. The attenuated profile is further broken up into two different components, the direct SBP and the scattered-in SBP, which are shown in orange and purple, respectively.}

\label{fig:sersic_4_u_fluxes}
\end{figure}

In this Appendix, we explore how each of the individual effects commonly grouped under the umbrella term of ``dust attenuation'' contributes to the observed attenuated SBP. From the SKIRT output, we can separate the observed attenuated flux into different components, which are plotted in Fig. \ref{fig:sersic_4_u_fluxes} for an example galaxy. 
\par
We find that the direct component, consisting of the light emitted along the line of sight that is neither absorbed nor scattered away, is the most relevant component in the observed attenuated SBP. Both absorption and scattering are more prevalent in the center of the galaxy, where the dust density is highest. However, we find that the direct SBP is shallower than the full attenuated SBP, with a fitted Sérsic index $n=2.585_{-0.008}^{+0.02}$, representing an even larger deviation in relation to the transparent SBP.

\par
For a full picture, we must also take into account that a significant amount of light is scattered into the line of sight, contributing to the re-injection of flux into the observed SBP, particularly in the central regions of the galaxy. This is a crucial detail, as many traditional dust corrections try to account for the lost flux, while only incidentally describing this scattered-in component as a variation in the slope of a given attenuation curve. As such, a full 3D approach is necessary to capture and disentangle all the nuances inherent in the data, especially for galaxies with high dust density regions, as seen in Section \ref{sec:lum_profile}.

\section{Infrared color gradients}
\label{app:k-band}
In this Appendix, we present results related to the dust-induced $V-K$ gradients. When dealing with near-infrared bands such as $K$, dust emission can no longer be safely neglected. As such, the simulations in this section were performed using the full ``DustEmission'' SKIRT mode. In addition, the \cite{Zubko_2004} dust mix properties are not properly defined over the full wavelength range required for these simulations. For this reason, we used a \cite{Draine_2007} dust mix instead.
\par
\begin{figure}[!h]
\centering
	\includegraphics[width=0.5\textwidth]{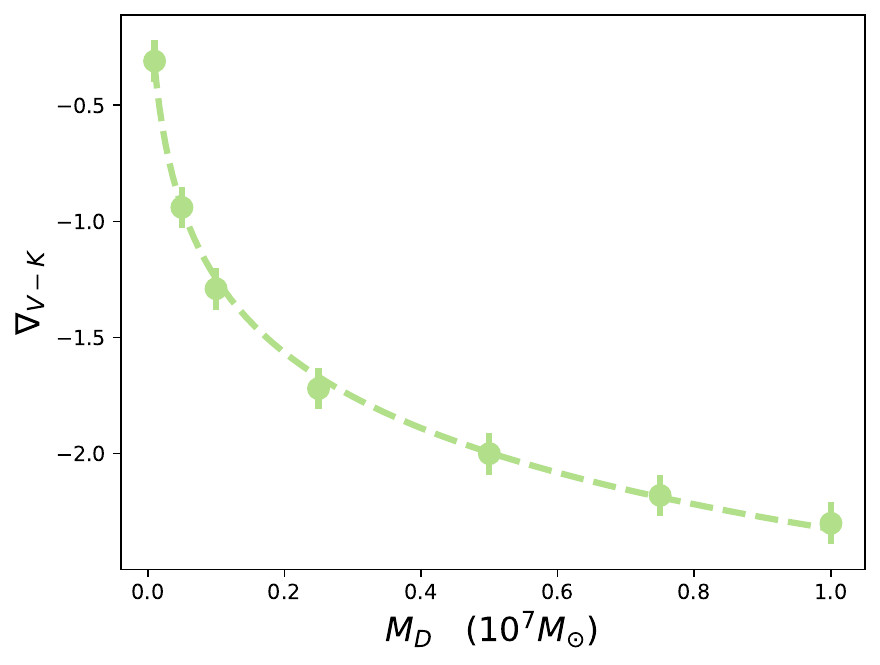}
	\caption{$V-K$ color gradients as a function of the dust mass $M_D$ for spherically-symmetric simulated galaxies with spatially invariant dust-to-stellar ration and a Sérsic profile with $n_{\textrm{sim}}=2$. The dashed lines correspond to exponential fits of the data.}

\label{fig:v-k}
\end{figure}
\begin{figure}[!h]
\centering
	\includegraphics[width=0.5\textwidth]{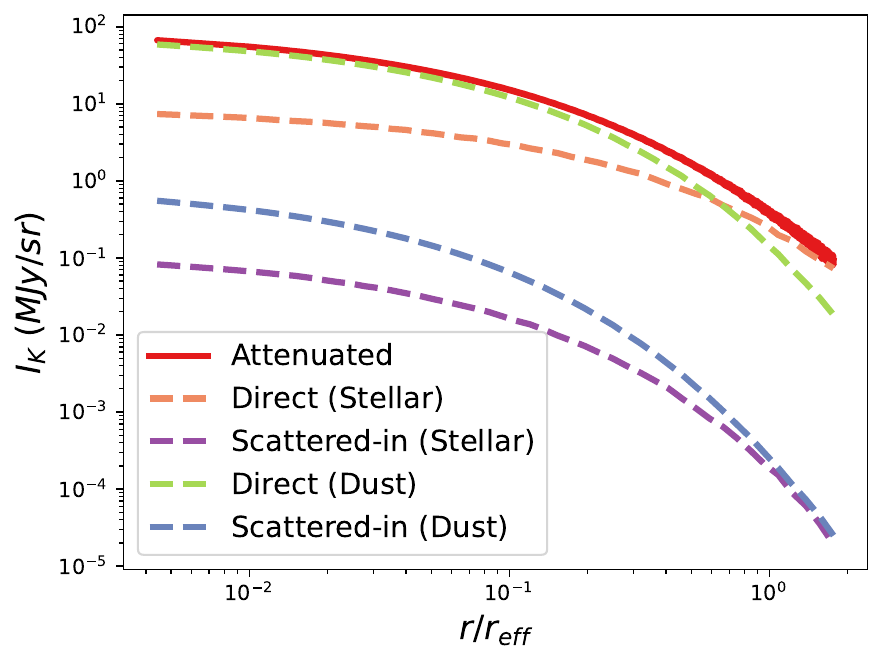}
	\caption{SBPs for a simulated Sérsic galaxy with $n_{\textrm{sim}}=2$, $r_{\textrm{eff,sim}}=4000$ pc and $M_{\mathrm{D}}=10^6 M_{\odot}$. The attenuated profile is shown in red, along with its underlying components: direct stellar flux (orange), scattered-in stellar flux (purple), direct dust emission flux (green), and scattered-in dust emission flux (blue).}

\label{fig:flux_k}
\end{figure}

The simulated $V-K$ gradients recovered for different dust masses are plotted in Fig. \ref{fig:v-k}. For illustration purposes, we limited these new simulations to the $n_{\textrm{sim}}=2$ case. A best-fit exponential curve is also shown. We find that these gradients reach the same order of magnitude as observed $V-K$ gradients (which can be as high as $\nabla_{V-K}= -0.6$ mag/dex \citep[e.g.,][]{Peletier90b}) for dust masses lower than $5\times10^5 M_{\odot}$. This means that, as expected, dust-induced color gradients are even more impactful in the optical-NIR than in the optical. As shown in Fig. \ref{fig:flux_k}, the recovered gradients, as well as the underlying SED, are primarily driven by the direct dust emission component, the strength of which depends on the radial dust density.

\par

\par

\section{Impacts of the stellar population on dust-induced color gradients}
\label{app:gal_age}
As discussed in Section \ref{sec:color_grad}, the color gradients obtained from our SKIRT radiative transfer simulations are completely dust-induced and do not depend on the SSP used. This is because, since a common SSP is assumed across the entire galaxy, any existing color from different stellar ages or metallicities is factored out from the gradient.
\par
\begin{figure}[!h]
	\centering
	\includegraphics[width=\textwidth]{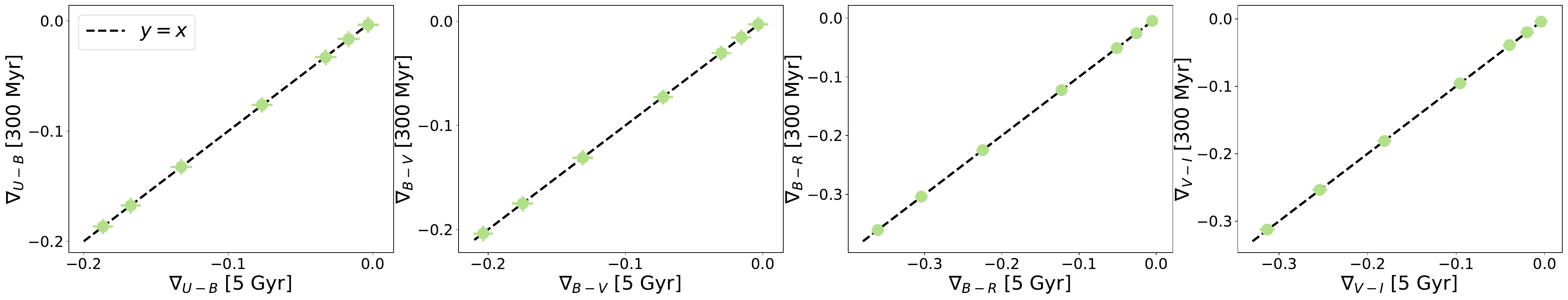}
	\caption{Color gradients for two identical simulated galaxies with different stellar population ages. Gradients for $U-B$, $B-V$, $B-R$ and $V-I$ are shown. Different points on the plot correspond to different dust masses, from $10^5 M_{\odot}$ to $10^7 M_{\odot}$. The dashed lines correspond to $y=x$.}

\label{fig:grads_300}
\end{figure}

To better highlight this property, we repeated our simulations using a slightly different SED model, roughly corresponding to a post-starburst galaxy. For this, we used a SSP with an age of $300$ Myr, keeping all other parameters the same as described in Section \ref{sec:methods_sim}. For illustration purposes, we limited these new simulations to the $n_{\textrm{sim}}=2$ case. The comparison between the color gradients recovered for the post-starbust galaxy simulations and the results for the ETG model described in \ref{sec:color_grad} is shown in Fig. \ref{fig:grads_300}. Different points on the plot correspond to different simulated dust masses, from $10^5 M_{\odot}$ to $10^7 M_{\odot}$.

%Appendix
\par
The recovered color gradients depend only on the star-to-dust geometry and the dust mass and are thus unaffected by the choice of SSP. However, we note that, for more complex and realistic galaxies, the mixing of different SSPs with different spatial distributions will, in all likelihood, complicate this picture. This is because the dust- and population-induced color gradients are not fully independent, being rather deeply intertwined (see \cite{Duarte_2025}). This means that, given a set of galaxy observations, it is not easy to disentangle both effects without comprehensive 3D radiative transfer modeling.

\end{appendix}

\end{document}